\documentclass{article}

\usepackage[letterpaper,top=2cm,bottom=2cm,left=3cm,right=3cm,marginparwidth=1.75cm]{geometry}

\usepackage{amsmath}
\usepackage{graphicx}
\usepackage{graphicx}
\usepackage{dcolumn}
\usepackage{bm}

\usepackage{graphicx} 
\usepackage{array}
\usepackage{multirow}
\usepackage{booktabs}
\usepackage{rotating}
\usepackage{tabu}
\usepackage{relsize}
\usepackage{amssymb}
\usepackage{pifont}
\usepackage{bbding}
\usepackage{starfont}
\usepackage{wasysym}
\usepackage{eqnarray}
\usepackage{tablefootnote}
\usepackage{ragged2e}
\usepackage[numbers]{natbib}
\usepackage{xcolor}
\usepackage{amsmath}

\definecolor{lightred}{RGB}{255, 128, 128}
\definecolor{lightgreen}{RGB}{128, 255, 128}
\definecolor{lightblue}{RGB}{128, 128, 255}
\definecolor{lightcyan}{RGB}{128, 255, 255}
\definecolor{lightmagenta}{RGB}{255, 128, 255}
\definecolor{lightyellow}{RGB}{255, 255, 128}
\definecolor{peach}{RGB}{255, 218, 185}
\definecolor{lavender}{RGB}{230, 230, 250}
\definecolor{aqua}{RGB}{127, 255, 212}
\definecolor{palepink}{RGB}{255, 182, 193}
\definecolor{tan}{RGB}{210, 180, 140}
\definecolor{silver}{RGB}{192, 192, 192}
\definecolor{mintgreen}{RGB}{152, 251, 152}
\definecolor{peachpuff}{RGB}{255, 218, 185}
\definecolor{lightgray}{RGB}{211, 211, 211}
\definecolor{darkblue}{RGB}{51,51,204}
\definecolor{burgundy}{RGB}{179, 51, 64}
\definecolor{color1}{RGB}{229, 25, 127}
\definecolor{color2}{RGB}{127, 25, 229}
\usepackage{tikz}

\title{Dynamics of fluid-driven fractures across material heterogeneities}
\author{Sri Savya Tanikella$^1$, Marie C. Sigallon$^{1,2}$, Emilie Dressaire$^{1}$\\
1. University of California Santa Barbara\\
2. Ecole Polytechnique, CNRS, Institut Polytechnique de Paris}

\begin{document}
\maketitle

\begin{abstract}
Fracture propagation is highly sensitive to the conditions at the crack tip. In heterogeneous materials, microscale obstacles can cause propagation instabilities. Macroscopic heterogeneities modify the stress field over scales larger than the tip region. Here, we experimentally investigate the propagation of fluid-driven fractures through multilayered materials. We focus on analyzing fracture profiles formed upon injection of a low-viscosity fluid into a two-layer hydrogel block. Experimental observations highlight the influence of the originating layer on fracture dynamics. Fractures that form in the softer layer are confined, with no penetration in the stiffer layer. Conversely, fractures initiated within the stiffer layer experience rapid fluid transfer into the softer layer when reaching the interface. We report the propagation dynamics and show that they are controlled by the toughness contrast between neighboring layers, which drives fluid flow. We model the coupling between elastic deformation, material toughness, and volume conservation. After a short transient regime, scaling arguments capture the dependence of the fracture geometry on material properties, injection parameters, and time. These results show that stiffness contrast can accelerate fracture propagation and demonstrate the importance of macroscopic scale heterogeneities on fracture dynamics. These results have implications for climate mitigation strategies involving the storage of heat and carbon dioxide in stratified underground rock formations.
\end{abstract}

\section{Introduction}
Fractures propagate when the elastic stress conditions at the crack tip are favorable to the formation of new surfaces \cite{Lawn1993}. The advance of the crack front is, therefore, controlled by local properties, including material composition, stress, and local heterogeneities. When the energy available within a material reaches the energy required to create new fracture surfaces, the crack propagates through an intermittent process that starts with a localized nucleation, which moves the tip forward \cite{Schmittbuhl1997, Wang2020, Cochard2024}. At the nucleation side, the high curvature of the crack front triggers rapid lateral growth. The irregular propagation is exacerbated by micro-heterogeneities that can pin the fracture front \cite{Gao1989, Ramanathan1997, Chopin2018, Steinhardt2022}. 
Heterogeneities whose characteristic length scale is larger than the tip region have also been reported to affect fracture growth \cite{Chopin2011, Xia2012, Roch2023}. Such heterogeneities are found in natural and engineered substrates, including rocks, adhesives, and composite materials. In heterogeneous substrates, fluid-driven fractures can be diverted by a plane of lower toughness, such as the interface between two material layers \cite{Tang2006, Santillan2017, Luo2018, Saez2022}. As the fracture tip propagates through a single material at any given time, understanding how macroscopic heterogeneities affect propagation requires coupling stress distribution and local material rupture conditions.

This study focuses on the interaction of a penny-shaped fracture with a material discontinuity in a brittle elastic medium with no stress contrast. Fluid-driven fractures have previously demonstrated that non-local effects, such as viscous dissipation \cite{Lai2015, Detournay2016, Tanikella2022, Tanikella2023} and in-situ stress distribution can affect the fracture propagation \cite{Yew2014}. Yet the role of heterogeneous material properties on a macroscopic scale and the corresponding fracture dynamics remain to be studied. We investigate fracture propagation when a fluid is injected into a hydrogel block composed of two layers that exhibit different stiffness. Since the material toughness governs the propagation, the stiffness contrast determines the fate of the fracture at the interface between the two layers. We are particularly interested in the fracture dynamics when the tip crosses from a stiff into a soft layer, as the macroscopic heterogeneity influences the propagation rate over length scales larger than the crack tip. The theoretical model of the propagation dynamics relies on scaling arguments to couple volume conservation, elastic stress, and fracture propagation criterion. The model yields a good agreement with experimental results. 

\section{Experimental methods}
\begin{figure}
\centering
\includegraphics[scale = 0.4]{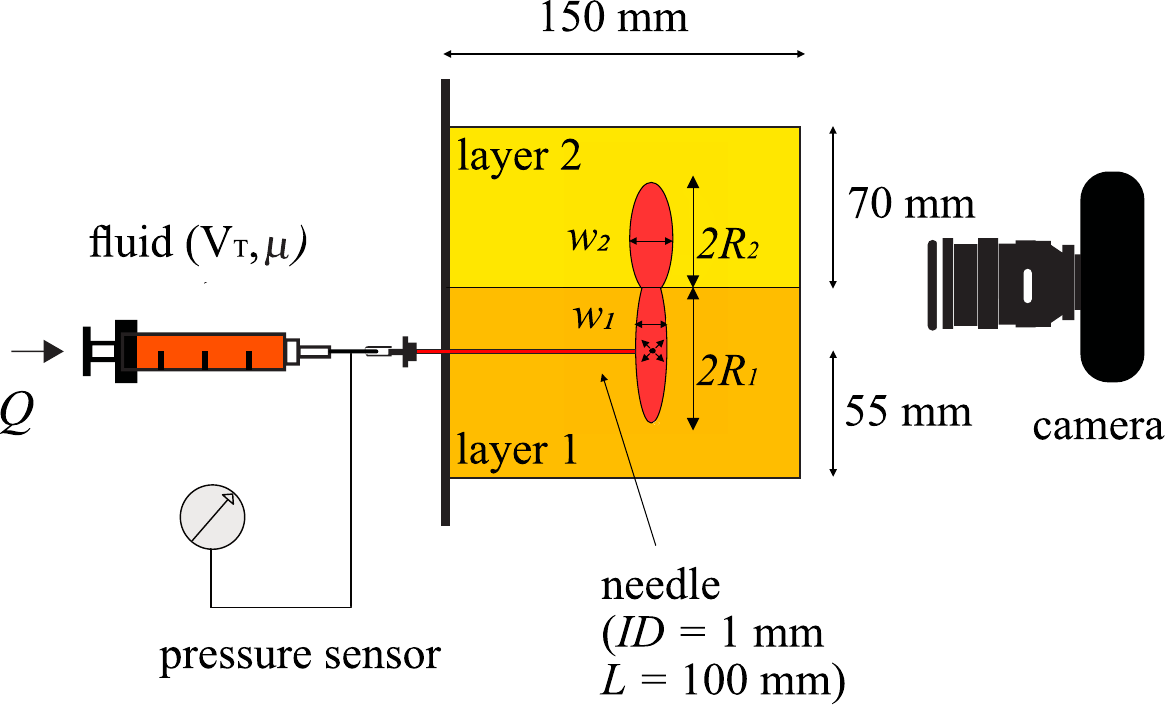}    %
\caption{Schematic of the experimental set-up.}\label{fig:SUP}
\end{figure}

The experimental system allows the study of planar fractures growing across a multilayered substrate under controlled loading conditions. We form the fracture upon injection of liquid in hydrogel. We use a substrate composed of two horizontal gelatin layers of comparable height in an acrylic cube of dimensions equal to 150 mm, as shown in Fig. \ref{fig:SUP} \cite{Giuseppe2009, Kavanagh2013, Lai2015, Tanikella2022, Tanikella2023}. The bottom and top layers are labeled 1 and 2, and their Young's moduli are $E_1$ and $E_2$, respectively. The relative masses of ultra-pure water and gelatin powder (Gelatin type A; Sigma-Aldrich, USA) to achieve the Young's modulus $E_1$ \citep{Tanikella2022} are combined and heated at 60$^o$C. The mixture is poured into the cube around a needle of inner diameter \textit{ID} = 1 mm \cite{Bunger2013}. 
We use a small plastic washer at the tip of the blunt injection needle to ensure the fracture propagates in the vertical plane, in the focal plane of the camera lens. The gelatin for the second layer is prepared using a different amount of gelatin powder to obtain Young's modulus $E_2$. To avoid softening the first layer while ensuring adhesion of the second layer, the gelatin mixture for layer 2 is cooled from 60$^o$C down to 40 $^o$C before pouring on top of layer 1. The hydrogel block then sets over 24 hours before fracturing \citep{Ham2020}. 
To form the fracture, silicone oil is injected into the block at the tip of a needle at a constant flow rate $Q$ with a syringe pump. As the volume of fluid increases, the fracture grows and reaches the interface between the two layers for $V=V_0$ and propagates until the injection stops at $V=V_T$. We vary the stiffness of the two layers, fluid viscosity, and injection parameters (flow rate and total volume), as detailed in Table \ref{tab:my-exp}. 
These parameters ensure the fracture propagation occurs within the toughness-dominated regime for both layers \cite{Savitski2002, Lai2015, OKeeffe2018a}: the viscous dissipation associated with the flow in the fracture is negligible. In the toughness-dominated regime, the elastic deformation of fracture surfaces leads to its radial expansion. The material properties are measured in the laboratory as described in \cite{Tanikella2022} and the measurement errors are estimated to $10\%$ for Young's modulus $E$, $10\%$ for the fracture toughness $K$, $0.35\%$ for the flow rate $Q$ and, $1\%$ for the viscosity $\mu$.

As the fracture forms and grows, we record images with a DSLR camera; the oil is dyed with Sudan red to improve the contrast and allow accurate area measurement in the plane of propagation. The fluid pressure is measured with a membrane sensor (ATO-PRES-P350) placed on the oil tubing connecting the syringe and the needle. We also record the fracture aperture using the light attenuation method \cite{Bunger2006, Tanikella2022} with Sudan red at 0.025 g.L$^{-1}$: the light absorbance at $\lambda = 520$ nm varies linearly as a function of the fracture aperture. We obtain aperture values across the fracture after calibration in a custom-made acrylic wedge \citep{Tanikella2022}.

\begin{figure}
\centering
\includegraphics[scale = 0.85]{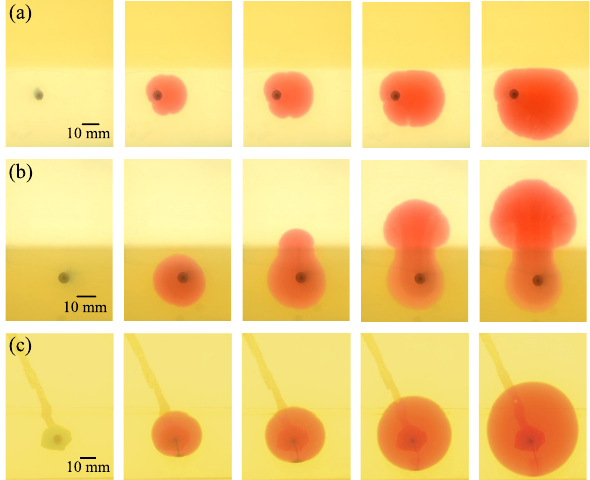}    %
\caption{Times series of the fracture propagation for (a) experiments 1, (b) experiment 2, and (c) experiment 3 at times t = [0, 120, 180, 330, 590] s, from left to right with $t = 0$ the onset of the fracture in layer 1. The layers have Young's moduli $E_{\mbox{soft}} = 15$ kPa (clear yellow) and $E_{\mbox{stiff}} = 30$ kPa (dark yellow). In experiment 3, the layers have the same Young's modulus $E_{\mbox{stiff}} = 30$ kPa.}\label{fig:OBS}
\end{figure}

\section{Results}
\subsection{Phenomenology}
As oil is injected into the gelatin, a fracture forms around the needle tip in layer 1 and propagates axisymmetrically in the vertical plane. The penny-shaped fracture expands towards the interface between layers 1 and 2, as shown in Fig. \ref{fig:OBS}.
\begin{table}[h]
\centering
\caption{List of experiments: symbols and parameters}
\begin{tabular}{clccccccc}
\multicolumn{1}{l}{} &
   &
  \multicolumn{1}{c}{$E_{1}$} &
  \multicolumn{1}{c}{$E_{2}$} &
  \multicolumn{1}{c}{$V_T$} &
  \multicolumn{1}{c}{$\mu$ } & 
  \multicolumn{1}{c}{$Q$ } &
   \multicolumn{1}{c}{\boldmath$V_0$\unboldmath} &
   \multicolumn{1}{c}{\boldmath${R_{1,f}}$\unboldmath}\\
  
\multicolumn{1}{l}{Exp.} & &
  (kPa) &
    (kPa) &
    {(ml)} &
(Pa.s) &
  (ml/min) &
  (ml) &
  (m)\\  
    \hline
\multicolumn{1}{l}{} &
  \multicolumn{1}{l}{} &
  \multicolumn{1}{l}{} &
  \multicolumn{1}{l}{} &
  \multicolumn{1}{l}{} &
  \multicolumn{1}{l}{} &
  \multicolumn{1}{l}{} &
  \multicolumn{1}{l}{} \\
1$^{\dag}$  & $\color{burgundy} \blacksquare \hspace{-2.65mm}\color{black} \square$       & 15 & 30 &  12 &  $0.02$ & 1  & - & -\\
2*$^{\dag}$  &   $\color{silver} {\mathlarger{\blacklozenge}}  \hspace{-3.4 mm} \color{black}$ {\raisebox{-0.0pt}{{$\lozenge$}}}           & 30  & 15 &  12 & $0.02$ & 1 & 2.77 &0.016\\
3  &   \color{lightyellow}{\ding{72}}\hspace{-4.2 mm} \color{black}{\raisebox{-2pt}{{\FiveStarOpen}}}           & 30  & 30 &  10 & $0.02$ & 1 & - & -\\
4  &   $\color{lightmagenta} {\mathlarger{\mathlarger{\blacktriangle}}}  \hspace{-4.2 mm} \color{black}$ {\raisebox{-0.5pt}{{$\triangle$}}}           & 15 & 30 &  10 & $0.01$ & 0.5 & - & - \\
5 &  \textcolor{lavender}{\rotatebox[origin=c]{-90}{\large$\blacktriangle$}} \hspace{-5.5mm} \raisebox{7.5pt}{\rotatebox{-90}{$\bigtriangleup$}} & 15 &  60  &  10 & $0.02$ & 1 & - & - \\
6*  & $\color{mintgreen}{\mathlarger{\mathlarger{\mathlarger{\bullet}}}} \hspace{-2.6 mm} \color{black} {\mathlarger{\mathlarger{\mathlarger{\circ}}}}$      & 30 & 60 & 10 & $0.02$ & 0.5 & - & - \\
7  &   $\color{lightblue} {\mathlarger{\blacklozenge}}  \hspace{-3.4 mm} \color{black}$ {\raisebox{-0.0pt}{{$\lozenge$}}}           & 30   & 15  &  12 & $0.1$ & 0.5 & 3.85 &0.018\\
8  &  \textcolor{lightcyan}{\rotatebox[origin=c]{180}{\large$\blacktriangle$}} \hspace{-5.55mm} \raisebox{7pt}{\rotatebox{180}{$\bigtriangleup$}}   & 60  &  15 &  10 & $0.02$ & 1 & 3.35 &0.016   \\
9 &  \textcolor{aqua}{\rotatebox[origin=c]{90}{\large$\blacktriangle$}} \hspace{-4.7mm} \raisebox{-1pt}{\rotatebox{90}{$\bigtriangleup$}}& 60 &  15  & 12 & $0.02$ & 0.5 & 3.68 &0.021  \\
10* & $\color{tan} \blacksquare \hspace{-2.65mm}\color{black} \square$  & 60  & 30 &  10 & $0.02$ & 0.5 & 4.04 &0.022  \\
11 & $\color{lightgreen} \blacksquare \hspace{-2.65mm}\color{black} \square$  & 30   & 15  & 10 & $0.01$ & 0.1 & 7.79&0.026 \\
12*  & $\color{peachpuff}{\mathlarger{\mathlarger{\mathlarger{\bullet}}}} \hspace{-2.5 mm} \color{black} {\mathlarger{\mathlarger{\mathlarger{\circ}}}}$      & 88 & 60 &  10 &  $0.02$ & 0.5  & 2.88 &0.018\\
13 & $\color{lightred}{\mathlarger{\mathlarger{\mathlarger{\bullet}}}} \hspace{-2.5 mm} \color{black} {\mathlarger{\mathlarger{\mathlarger{\circ}}}}$      & 30 & 15 &  10 &  $0.01$ & 0.5  & 5.3&0.023\\
2$^r$  &   $\color{darkblue} {\mathlarger{\blacklozenge}}  \hspace{-3.4 mm} \color{black}$ {\raisebox{-0.0pt}{{$\lozenge$}}}           & 30  & 15 &  12 & $0.02$ & 1 & 1.7 &0.012
\end{tabular}
\label{tab:my-exp}
\footnotesize{\begin{justify}
Pressure measurements for experiments marked with $^{*}$ and aperture measurements for experiments marked with $^{\dag}$. $\bf{V_0}$ and $\bf{R_{1,f}}$ are measured. In experiments $1$ and $4-6$, the fracture is confined in layer 1.
\end{justify}}
\end{table}

 To explore the role of the relative stiffness of the layers, we conduct three experiments with layers of Young's moduli $E=[15, 30]$ kPa (experiments 1 to 3 in Table \ref{tab:my-exp}). In experiment 1,  layer 1 is the softer layer, and the fracture growth is presented in Fig. \ref{fig:OBS}(a) and Movie S1. 
For experiment 2, layer 1 is the stiffer layer, and the results are in Fig. \ref{fig:OBS}(b) and Movie S2. Despite identical injection parameters ($Q = 1$ ml min$^{-1}$ and $\mu = 20$ mPa s), the fractures produced exhibit different geometry and propagation dynamics once they reach the interface. In experiment 1, the fracture continues to propagate in the soft layer, adopting an elongated geometry \cite{Zia2018}. 
In experiment 2, the fracture crosses over to the second and softer layer upon reaching the interface. The fracture initially adopts a half-disk shape in the soft layer. 
In comparison, when the two layers have the same Young's modulus, the fracture adopts a circular geometry (see Fig. \ref{fig:OBS}(c)) and Movie S3). This observation indicates that the contrast in stiffness between the layers is responsible for the different shape and dynamics and not the presence of an interface between two gelatin layers. 

To rationalize these results, we consider the pressure required to propagate the fracture in the different layers. The ratio depends on the material stiffness and can be approximated as $P_2/P_1 \propto \sqrt{\gamma_1G_1/\gamma_2G_2}$, where $\gamma$ is the fracture energy and $G$ the shear modulus \citep{Biot1983}. Since both layers are made of gelatin and have the same fracture energy $\gamma \approx $ 1 J m$^{-2}$, and Poisson's ratio $\nu \approx 0.5$ \cite{Menand2002}, the relation simplifies to $P_2/P_1 \propto \sqrt{E_1/E_2}$. Therefore, when the penny-shaped fracture reaches the interface between the two layers, it preferentially propagates in the layer with the lowest Young's modulus at the lowest propagation pressure \cite{Luo2018, Zia2018}. This is consistent with our observations: a fracture that originates in the softer layer remains confined in the softer layer, and a fracture that forms in a stiffer layer crosses the interface and propagates in the softer neighboring layer.

\subsection{Propagation in a homogeneous substrate}
We first consider the propagation of the fracture in the injection layer. When a penny-shaped fracture forms and grows in a homogeneous infinite material, the propagation dynamics are driven by elastic deformation and limited by the fracture opening or toughness of the material. In all experiments presented here, the viscous dissipation associated with the flow in the fracture is negligible. The fluid injection starts at $t_1=0$. At time $t_1$, the fracture volume equals $V = Qt_1$. We can model the propagation of the fracture using scaling arguments \cite{Savitski2002, Lai2016a, OKeeffe2018a}. The average radius of the fracture is noted $R_1$. The aperture of the axisymmetric fracture is maximum at the injection point where $w(r=0)=w_1$. The elastic pressure that drives the fracture propagation is $P_1$. Using those parameters, the fracture volume scales as:
\begin{figure}
\centering
\includegraphics[width = \columnwidth]{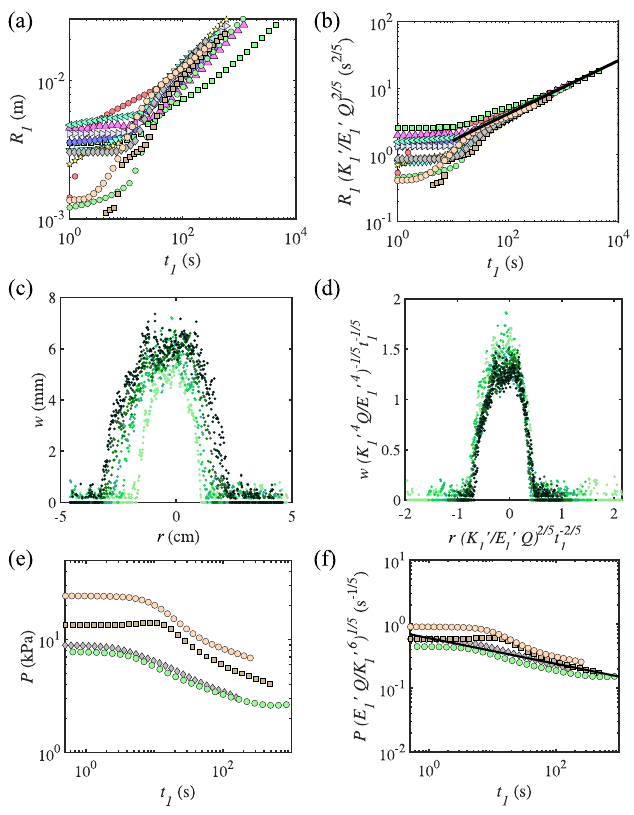}
\caption{Fracture dynamics in layer 1. (a) The time dependence of the radius is reported for all experiments and (b) rescaled following Eq. \ref{EqR1}. The black line represents the best-fit line with a slope of 1. (c) The radial dependence of the fracture aperture is measured at t = [100, 200, 300, 400, 500, 600, 700] s for experiment 1 and (d) the fracture profiles are rescaled using Eq. \ref{Eqw1}. (e) The pressure is measured over time for experiments 2, 6, 10, and 12. (f) The pressure data are rescaled with Eq. \ref{EqP1}. The black line represents the best-fit line with a slope of -1/5.}\label{fig:layer1}
\end{figure}
\begin{equation}
V \approx R_1^2w_1 \label{EqV1}
\end{equation}
Assuming there is no fluid lag at the low injection flow rates considered here \cite{Garagash2000}, the theory of Linear Elastic Fracture Mechanics (LEFM) holds. The fracture propagates in mobile equilibrium such that the elastic pressure is
\begin{equation}
P_1 \approx E'_1w_1/R_1 \label{EqP1}
\end{equation}
where $E_1' = E_1/(1-\nu^2)$ and $\nu \approx 1/2$ is the Poisson's ratio of gelatin. The stress intensity factor at the fracture tip should equal the material toughness for the fracture to grow. The fracture propagation criterion yields
\begin{equation}
K_1' \approx P_1\sqrt{R_1} = 8\sqrt{E_1'\gamma/\pi} \label{EqK1}
\end{equation}
where $\gamma = 1$ J m$^{-2}$ is the fracture energy for gelatin \cite{Rice1968, Okeeffe2018b, Menand2002}. These three equations yield scaling expressions governing $R_1$, $w_1$, and $P_1$ \citep{Savitski2002,Detournay2004},
 \begin{eqnarray}
     R_1 &\approx& \left( \frac{E_1'}{K_1'}\right)^{2/5}V^{2/5} \label{EqR1}\\
     w_1 &\approx& \left( \frac{K_1'}{E_1'}\right)^{4/5}V^{1/5} \label{Eqw1}\\
 P_1 &\approx& \left( \frac{K_1'^6}{E_1'}\right)^{1/5}V^{-1/5} \label{EqP1}
 \end{eqnarray}

In the toughness-dominated regime, the propagation dynamics depend on the material properties and flow rate at which the liquid is injected $Q$, as $V= Q\,t_1$. The radius measurements shown in Fig. \ref{fig:layer1}(a-b) demonstrate that the fracture growth in layer 1 is well-captured by Eq. \ref{EqR1}. The best-fitting power law for the experimental data is described by $R_1 = \alpha_1({E_1'}/{K_1'})^{2/5}(Qt)^{2/5}$, with $\alpha_1 = 0.65$, a value which is in agreement with previous studies \citep{Savitski2002, Lai2016a, OKeeffe2018a}. The fracture aperture is measured using light absorbance and plotted as a function of the radial position in Fig. \ref{fig:layer1}(c): the fracture increases in diameter and aperture during the injection. Rescaling both parameters using Eqs. \ref{EqR1} and \ref{Eqw1}, the fracture profiles collapse on a single curve, which is consistent with the self-similar geometry of the penny-shaped fracture (see Fig. \ref{fig:layer1}(d)). The fluid pressure can also be rescaled using Eq. \ref{EqP1} as shown in Fig. \ref{fig:layer1}(e-f). 
It is worth noting that the results presented in Fig. 3 include data collected when after the fracture has reached the interface between a soft and a stiff layer, and its growth is confined to the soft injection layer. The shape of the fracture becomes elongated. Yet, the time dependence of all three parameters follows the scaling behavior derived for an infinite substrate and an axisymmetric fracture, illustrating the robust growth dynamics of the self-similar penny-shaped fractures.
\begin{figure}[t]
\centering
\includegraphics[width=1\linewidth]{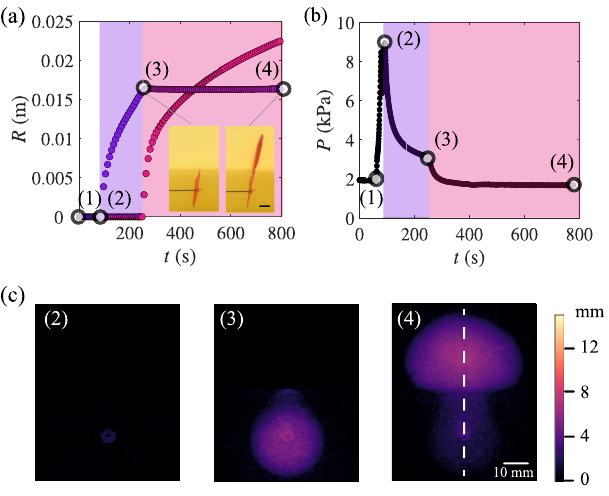}
\caption{Fracture dynamics in experiment 2: (a) Effective sub-fracture radius in layers 1 $\color{color2}{\mathlarger{\mathlarger{\mathlarger{\bullet}}}} \hspace{-2.1 mm} \color{black} {\mathlarger{\mathlarger{\mathlarger{\circ}}}}$ and 2 $\color{color1}{\mathlarger{\mathlarger{\mathlarger{\bullet}}}} \hspace{-2.1 mm} \color{black} {\mathlarger{\mathlarger{\mathlarger{\circ}}}}$. (b) Fluid pressure. (c) Fracture aperture. Fracture profiles are measured along the white dashed line. We note (1) the start of the injection, (2) the initiation of the fracture, (3) the crossing of the interface between layers 1 and 2, and (4) the end of the injection.}
\label{fig:prop}
\end{figure}

\subsection{Propagation through a layered substrate}
\subsubsection{Experimental results}
We now focus on the dynamics of a fracture that grows from a stiff into a soft layer, as observed in experiment 2. After the fracture crosses the interface, it comprises two sub-fractures: sub-fracture 1 in layer 1 and sub-fracture 2 in layer 2. We measure the areas of the sub-fractures and estimate the effective radii plotted in {Fig. \ref{fig:prop}(a)}. Before the fracture reaches the interface (at point 3), its radius increases in layer 1. Upon crossing the interface, sub-fracture 1 stops expanding, while the radius of sub-fracture 2 increases rapidly. We record the fluid pressure values, starting when the syringe pump is turned on, as shown in {Fig. \ref{fig:prop}(b)}. At the start of the injection (point 1), the fluid pressure increases until it reaches a critical value necessary to initiate the crack \cite{Lecampion2017}. Once the fracture forms (point 2), there is a rapid pressure release, and the propagation becomes quasi-steady in layer 1, following the scaling in Eq. 
\ref{EqP1}. The pressure drops rapidly when the fracture reaches the interface (point 3). After this transient regime, the pressure evolution becomes quasi-steady again until the end of the injection (point 4). Using light absorbance, we measure the aperture of the fracture, which is represented using a heat map in {Fig. \ref{fig:prop}(c)}. As the fracture propagates through layer 1, its aperture increases. When the fracture crosses the interface, the aperture of sub-fracture 1 decreases, whereas the aperture of sub-fracture 2 increases. 

These observations show that the coupling between the two sub-fractures controls the propagation of the fracture when it crosses an interface. Indeed, the collapse of sub-fracture 1 fuels the rapid growth of sub-fracture 2. This fluid transfer is driven by the larger elastic stresses in the stiffer layer 1. Based on our measurements, the radius of sub-fracture 1 remains constant, equal to $R_{1,f}$, while the maximum aperture $w_1(t)$ decreases over time. As the fracture crosses the interface, the pressure decreases from the value set by the toughness of layer 1, necessary to grow the fracture in layer 1, to the value set by the toughness of layer 2, as the fracture now only grows in layer 2. After a short transient regime, the pressure can be considered uniform in the two sub-fractures, with $P \approx P_1 \approx P_2$.   
\begin{figure}
\centering
\includegraphics[width = 0.85\columnwidth]{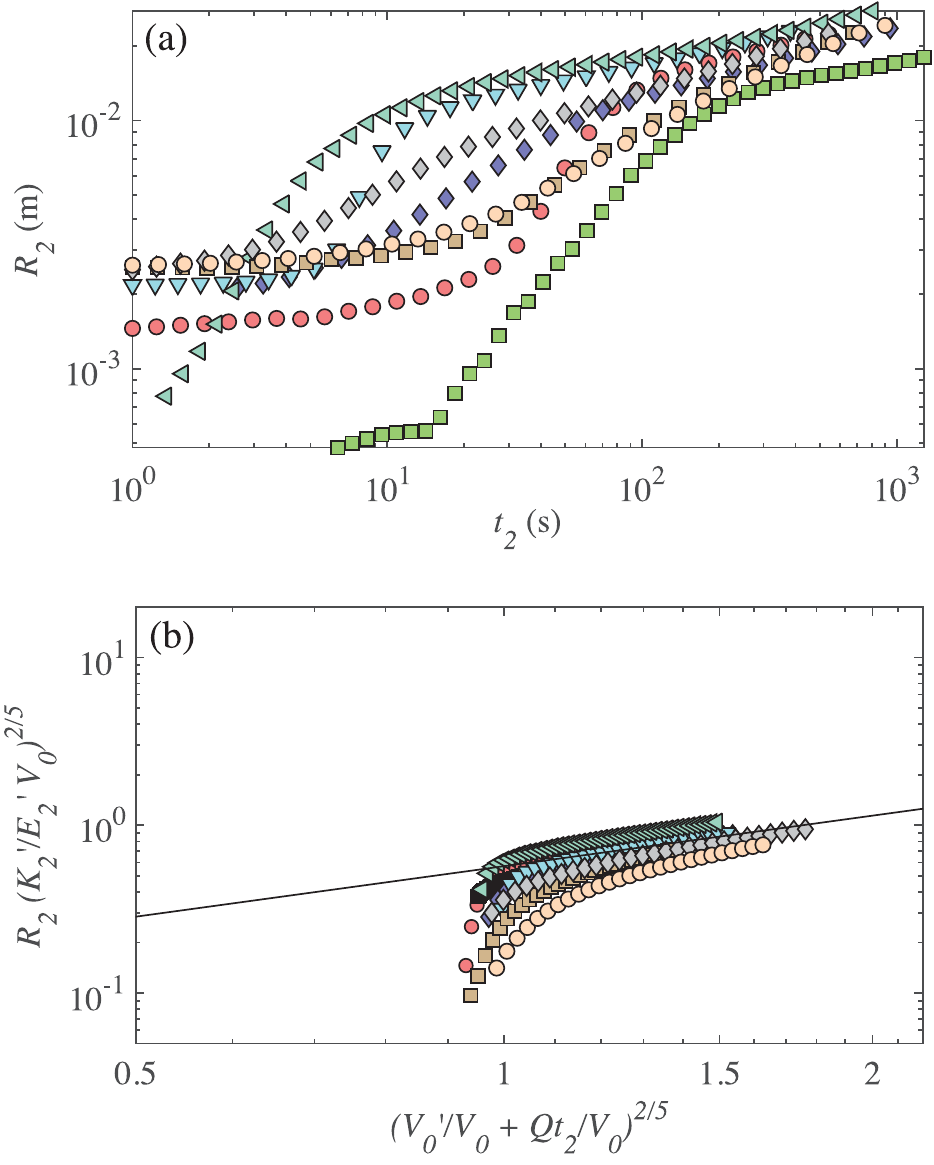}
\caption{Fracture radius in layer 2. (a) Effective radius for experiments 2 and 7-13 and (b) corresponding scaled effective radius using Eq. \ref{EqR2}. The black line represents the best-fit line with a slope of 1.}\label{fig:radiuslayer2}
\end{figure}
\subsubsection{Theoretical model}
To understand how the elastic relaxation of sub-fracture 1 is coupled with the propagation of sub-fracture 2, we use scaling arguments and derive expressions for the fracture parameters, i.e., the effective radius $R_2$ and maximum aperture $w_2$ in layer 2, maximum aperture $w_1$ in layer 1, and pressure $P$. 
We note $V_0$ the fracture volume when it reaches the interface. After crossing, the fracture volume is $V = V_0 + Qt_2$ where $t_2 = 0$ when the fracture reaches the interface and 
\begin{equation}
    V \approx R_{1,f}^2 \, w_1 + R_2^2 \, w_2 \label{EqV}. 
\end{equation}

The elastic pressure scales as $P_1 \approx E'_1w_1/R_{1,f}$ and $P_2 \approx E'_2w_2/R_2$ with $P \approx P_1 \approx P_2$ \citep{Savitski2002,Detournay2004}. We can approximate the fracture aperture in both layers as follows:
\begin{align}
w_1 & \approx \frac{P R_{1,f}}{E'_1},  \label{Eqw1b} \\
\mbox{and} \; \; w_2 & \approx \frac{P R_2}{E'_2} \label{Eqw2}
\end{align}

Assuming there is no fluid lag and the theory of Linear Elastic Fracture Mechanics (LEFM) holds, the material properties of layer 2 set the propagation criterion \cite{Garagash2000}: 
\begin{equation}
K'_2  = 8\sqrt{E_2'\gamma/\pi} \approx P\sqrt{R_2}. \label{EqK2}
\end{equation}
\begin{figure}[h]
\centering
\includegraphics[width = \columnwidth]{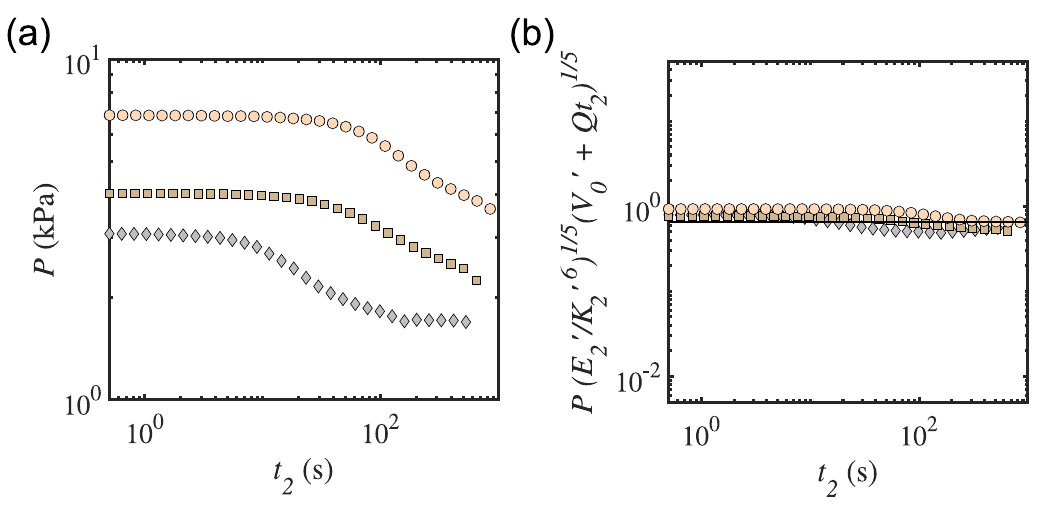}
\caption{Fluid pressure for experiments 2, 6, 10, and 12 in layer 2: (a) pressure as a function of time, and (b) rescaled pressure data using Eqs. (1) and (4).}\label{fig:pressurelayer2}
\end{figure} 
To predict the evolution of $R_2$, $w_2$, $w_1$, and $P$, the system of equations \ref{EqV}-\ref{EqK2} needs to be solved. The coupled equations yield a 6th-order polynomial equation whose exact roots are not conducive to capturing the physics of the fracture propagation. Consequently, we use the scaling expressions derived for an infinite uniform material to get approximate closed-form expressions for the fracture parameters. 

We hypothesize that the expressions for the radius, aperture, and pressure of sub-fracture 2 have the same form as the scaling relations for a fracture growing in a uniform infinite material. In an infinite material, the volume of the fracture is equal to the injected volume. Here, the volume of sub-fracture 2 is the sum of the volume transferred from sub-fracture 1, noted $V_0'$ and the injected volume $Qt_2$, which is assumed to flow entirely into sub-fracture 2 since sub-fracture 1 collapses:
\begin{equation}
    V_2 = V_0'+ Q\,t_2. \label{EqV2}
\end{equation}

The scaling relations \ref{EqR1}, \ref{Eqw1} and \ref{EqP1} become:
\begin{equation}
    R_2 \approx \frac{E_2'^{2/5}}{K_2'^{2/5}}(V_0' + Qt_2)^{2/5} \label{EqR2},
\end{equation}

\begin{equation}
    w_2 \approx \frac{K_2'^{4/5}}{E_2'^{4/5}}(V_0' + Qt_2)^{1/5} \label{Eqw2},
\end{equation}

\begin{equation}
    P \approx \frac{K_2'^{6/5}}{E_2'^{1/5}}(V_0' + Qt_2)^{-1/5} \label{EqP2}.
\end{equation}
\begin{figure*}
\centering
\includegraphics[scale = 0.85]{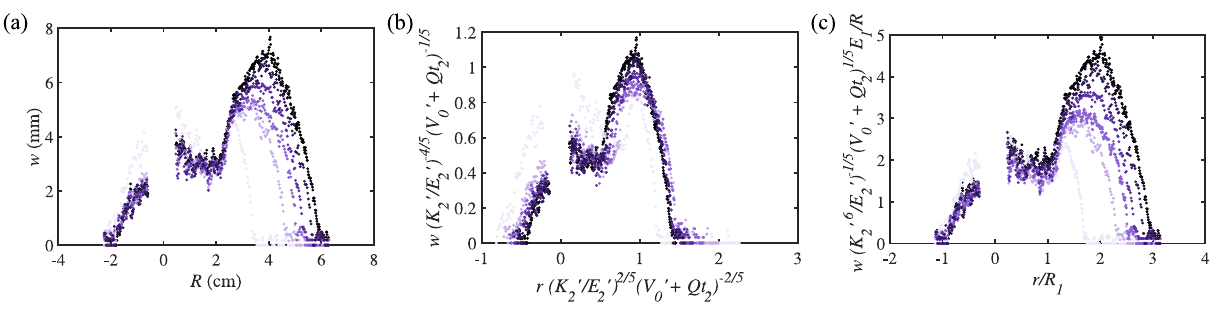}
\caption{Fracture profiles measured during experiment 2: (a) Aperture values $w$ at $t_2 = [21, 111, 201, 291, 381, 471]$ s, with time increasing from light to dark purple, (b) rescaled aperture profiles using Eqs. (2) and (3), for the propagation in layer 2, and (c) rescaled aperture profiles using $R_{1,f}$ and Eq. (5), for the collapse in layer 1.}\label{fig:thicknesslayer2}
\end{figure*}
Substituting Eq. \ref{EqP2} in Eq. \ref{Eqw1b}, we solve for $w_1$:
\begin{equation}
    w_1 \approx \frac{K_2'^{6/5}}{E_2'^{1/5}}(V_0' + Qt_2)^{-1/5}\frac{R_{1,f}}{E_1'}. \label{Eqw1c}
\end{equation}

The volume of the fluid $V_0'$ from layer 1 in layer 2 at time $t_2$ is 
\begin{equation}
    V_0' \approx V_0 - R_{1,f}^2\,w_1.\\ \label{EqwV0p}
\end{equation}
Substituting the expression for $w_1$ in Eq. \ref{Eqw1c}, we define the transferred volume:
\begin{equation}
  V_0'  \approx V_0 - \frac{K_2'^{6/5}}{E_2'^{1/5}}\frac{R_{1,f}^3}{E'_1} (V_0 + Qt_2)^{-1/5}.\label{EqwV0pb}
\end{equation}

The volume transferred $V_0'$ is an increasing function of time that converges toward $V_0$ at large times when the fluid initially in sub-fracture 1 is completely transferred into sub-fracture 2. In the limit of $t_2 = 0$, the transferred volume has a finite value $V_0'(t_2=0) = V_0 \left[1-(E_2/E_1)^{2/5}\right]$. This initial transferred volume equals 0 when the two layers have the same Young's modulus, and no fluid flow between the two layers is expected. The value of the initial transferred volume is finite for different layers as the model does not capture the rapid transient regime during which the pressure in the two layers becomes equal to $P_2$. The initial transferred volume increases as the ratio of Young's moduli increases, and more fluid flows into sub-fracture 2 during the transient regime. For small values of $t_2$, the transferred volume increases linearly with time. This results in the faster propagation of sub-fracture 2, with an increase in effective flow rate. The contribution of sub-fracture 1 to the growth of sub-fracture 2 decreases over time. 
\subsubsection{Discussion}
To compare the expressions derived with experimental results, we focus on the fracture dynamics once the tip crosses the interface between the two layers. For the range of parameters in Table \ref{tab:my-exp}, we measure the effective radius of sub-fracture 2; the results are presented in Fig. \ref{fig:radiuslayer2}(a). Initially, we observe a rapid increase in radius, characteristic of the swift transfer occurring between the two layers. Using Eqs. \ref{EqR2} and \ref{EqwV0pb} to define the dimensionless radius of sub-fracture 2 and the transferred volume, all datasets collapse on a master curve as shown in Fig. \ref{fig:radiuslayer2}(b). The best-fitting power law for this unified curve is $R_2 ({K_2'}/{E_2'\,V_0})^{2/5}= \alpha_2(V_0'/V_0+Qt_2/V_0)^{2/5}$, with $\alpha_2 = 0.57$. 
This prefactor is consistent with the values obtained in the toughness regime for an infinite homogeneous material \cite{OKeeffe2018a, Lai2016a, Tanikella2022}. 
Figure \ref{fig:pressurelayer2}(a) reports the liquid pressure once the fracture has advanced into layer 2. When rescaled, the pressure data converge to a constant, as depicted in Fig. \ref{fig:pressurelayer2}(b). The best-fit line is given by $P = \alpha_4 ({K_2'^{6}}/{E_2'})^{1/5}(V_0' + Qt_2)^{-1/5}$, where $\alpha_4 = 0.65$. The results validate the hypothesis that the fluid pressure is set by the toughness of layer 2. 
The shape of sub-fracture 2 is initially ellipsoidal as the fluid transfer from sub-fracture 1 occurs over a distributed interface region. The fracture apertures $w_1$ and $w_2$ are measured along a line normal to the interface and going through the injection point, as shown in Fig. \ref{fig:prop}(c). The results presented in Fig. \ref{fig:thicknesslayer2}(a) show that the aperture of sub-fracture 1 decreases as the aperture of sub-fracture 2 increases over time. The aperture of fracture 2 reaches values up to 7 mm.
We rescale the fracture profiles with the maximum radius $R_2$ and aperture $w_2$, using Eqs. \ref{EqR2} and \ref{Eqw2}. We present the results in Fig. \ref{fig:thicknesslayer2}(b): the profiles of sub-fracture 2 collapse, indicating that the fracture geometry is self-similar and the scaling for $R_2$ and $w_2$ agree with the measurements, despite the elongate shape of sub-fracture 2. If we rescale the fracture aperture using Eq. \ref{Eqw1c} for $w_1$ and a fixed radius, $R_{1,f}$, the profiles of sub-fracture 1 collapse onto each other, as shown in Fig. \ref{fig:thicknesslayer2}(c). We conclude that the sub-fractures exhibit self-similar profiles, which are robustly observed for planar or penny-shaped hydraulic fractures \cite{Lai2016b, Okeeffe2018b}. 
\section{Conclusion}
Our results demonstrate the long-range interactions between macroscopic material heterogeneities and fracture dynamics: the fracture can undergo enhanced or confined propagation because of differences in stiffness between two neighboring layers. Our results show that a fracture growing from a stiff into a soft layer experiences faster expansion dynamics due to fluid transfer from the stiff to the soft layer. The fracture geometry also depends on the growth history and relative stiffness of the layers encountered. The proposed model uses scaling arguments to describe fluid transfer from high to low elastic stress regions and fracture propagation. After a short transient regime, the model captures the geometric parameters of the fracture and the fluid pressure. The role of the fluid is a notable difference between the influence of micro-scale and macro-scale heterogeneities on fracture properties. For micro-heterogeneities, the deformation of the crack front around the obstacle is responsible for the micro-instabilities in the propagation. Macro-heterogeneities are responsible for inhomogeneous stress fields, which drive flow in a fluid-filled fracture. For a low-viscosity Newtonian liquid, the fluid transfer is controlled by the differences in elastic stresses between the two sub-fractures. More generally, the compressibility and rheology of the fluid are expected to influence the fracture growth. A description of fracture dynamics combining the effects of micro and macro-heterogeneities will result in a better understanding of the propagation in disordered media.

The authors acknowledge the support of the UCSB Academic Senate and thank Dr. Emre Turkoz for helpful conversations.

\bibliographystyle{abbrv}

\end{document}